\magnification\magstep1
\baselineskip = 0.7 true cm
\parskip=0.5 true cm
                           
  \def\sa{\vskip 0.30 true cm}
  \def\sb{\vskip 0.60 true cm}


\def\gr{\hbox{\bf R}}
\def\grt{\hbox{{\bf R}$^3$}}
\def\grn{\hbox{\bf N}}
\def\grz{\hbox{\bf Z}}

\def\r3{{\bf r}
         \kern-0.6em{\raise 0.65em \hbox{. \kern-0.5em{.} \kern-0.5em{.}} }}
\def\z3{z\kern-0.6em{\raise 0.65em \hbox{. \kern-0.5em{.} \kern-0.5em{.}} }}
\def\y3{y\kern-0.6em{\raise 0.65em \hbox{. \kern-0.5em{.} \kern-0.5em{.}} }}
\def\x3{x\kern-0.6em{\raise 0.65em \hbox{. \kern-0.5em{.} \kern-0.5em{.}} }}

\rightline{\bf LYCEN~9155}

\rightline{December 1991}

\sa
\sb
\sa

\centerline {\bf Classical Trajectories for Two Ring-Shaped Potentials}

\sa
\sb

\centerline {MAURICE KIBLER AND GEORGES-HENRI LAMOT} 

\sa

\centerline {Institut de Physique Nucl\'eaire de Lyon,} 
\centerline {IN2P3-CNRS et Universit\'e Claude Bernard,} 
\centerline {F-69622 Villeurbanne Cedex, France} 

\sa
\sa

\centerline {PAVEL WINTERNITZ}

\sa

\centerline {Centre de Recherches Math\'ematiques,} 
\centerline {Universit\'e de Montr\'eal, CP 6128-A,} 
\centerline {Montr\'eal, Qu\'ebec, Canada H3C 3J7} 

\sa
\sb

\baselineskip = 0.65 true cm

\sa
\sa
\sb
\sa
\sa
\sb
\sa
\sb
\sb

\leftskip  = 0.3 true cm
\rightskip = 0.3 true cm

\noindent This work 
[published in {\bf Int. J. Quantum Chem. 43, 625 (1992)}] 
has been achieved  in the framework of the 
France-Qu\'ebec  exchange programme   (project no 20~02~20~89). 
The kind hospitality extended to one of the authors (M.~K.) 
at the Centre de Recherches 
Math\'ematiques de l'Universit\'e de Montr\'eal on the occasion 
of several stays during the work on this project  is gratefully 
acknowledged.      The research of one of the authors (P.~W.) 
is partially supported by 
research grants from NSERC of Canada and FCAR of Qu\'ebec. Two 
of the authors (M.~K. and P.~W.) 
are indebted to Dr.~N.~W.~Evans from the Queen Mary 
College (London) for sending them preprints on his work and to 
Professor Ya.~A.~Smorodinsky for an interesting discussion. 

\leftskip  = 0 true cm
\rightskip = 0 true cm

\vfill\eject
\baselineskip = 0.74 true cm
\centerline {\bf Abstract}

\sa
\sb

The present paper deals with the classical 
trajectories for two super-integrable systems~: a system 
known in quantum chemistry as the Hartmann system and a system 
of potential use in quantum chemistry and nuclear physics. Both 
systems correspond to ring-shaped potentials. They 
admit two maximally super-integrable systems as limiting cases, 
viz, the  isotropic harmonic oscillator  system and the  Coulomb-Kepler 
system in three dimensions.        The planarity of the trajectories is 
studied in a systematic way.           In general, the trajectories are 
quasi-periodic rather than periodic.    A constraint condition allows to 
pass from quasi-periodic motions to periodic ones. When written 
in a quantum mechanical context,  this constraint condition leads to new 
accidental degeneracies for the two systems studied. 

\vfill\eject

\centerline {\bf 1. Introduction}
\bigskip

Three-dimensional 
potentials that are singular along curves have received a 
great deal of attention in the recent years. In particular, the 
coulombic ring-shaped potential $V_Q$ (see Eq.~(3.1)), 
revived in quantum chemistry by Hartmann and collaborators 
[1], and the oscillatory ring-shaped 
potential $U_Q$ (see Eq.~2.1)), systematically studied by 
Quesne [2], have been investigated from a quantum mechanical 
viewpoint by using various 
approaches (including Schr\"odinger, Feynman path integral, 
nonbijective canonical transformations, and Lie-like 
approaches). In this connection, we mention the works of 
Refs.~[3-9] on the potential $V_Q$. 
Furthermore, the potential $U_Q$ and various 
extensions of $U_Q$ and $V_Q$ 
have been worked out by means of algebraic, path integral, and Lie-like 
techniques [10-14]. Finally, the 
so-called ABC and ABO systems (which are companions of the $V_Q$ 
and $U_Q$ systems concerning an Aharonov-Bohm 
plus a Coulomb potential and an Aharonov-Bohm 
plus an oscillator potential, respectively) have been the object of 
several investigations in a nonrelativistic formulation 
[15-18] as well as in a relativistic one [19].

The $U_Q$ and $V_Q$ systems have been little examined from a 
classical viewpoint [7,20]. It is 
the aim of this paper to study the classical motion for a 
(charged) particle embedded in the potential $U_Q$ or $V_Q$. 
More specifically, we shall investigate the planarity, the 
periodicity and the semi-classical quantization of the bounded 
motions for the $U_Q$ and $V_Q$ systems.

Among the potentials in $n = 3$ dimensions, the potentials $U_Q$ 
and $V_Q$ have a special status in the sense that they correspond 
to super-integrable hamiltonian systems with $n + 1 = 4$ integrals 
of motion. The latter two systems occupy a position 
intermediate between integrable hamiltonian systems (with $n = 3$ 
constants) and maximally super-integrable hamiltonian systems 
(with $2n - 1 = 5$ constants). In addition, when the parameter 
$Q$ vanishes, the $U_Q$ and $V_Q$ systems become maximally 
super-integrable systems ; as a matter of 
fact, $U_{Q=0}$ corresponds to the oscillator system and 
      $V_{Q=0}$             to the Coulomb-Kepler system. Let 
us recall that, in $n$ dimensions ($n$ arbitrary), 
four maximally super-integrable hamiltonian systems 
are known~: the oscillator system, the Coulomb-Kepler system, 
the Calogero-Moser system, and a system introduced by Smorodinsky and 
co-workers [21,22] (see also Ref.~[23]). 

The $U_Q$ and $V_Q$ systems thus turn out to be two interesting 
laboratories among the super-integrable systems. As ring-shaped 
systems, they may play an important role in all situations where 
axial symmetry is relevant. For example, the $V_Q$ 
system is of interest for ring-shaped 
molecules like cyclic polyenes [1]. Further, the $U_Q$ system 
is of potential use in the study of (super-)deformed nuclei. 

The plan of the present paper is as follows. For each of the 
potentials $U_Q$ and $V_Q$ we study in Sections 2 and 3, 
respectively, the constants of motion, the classical 
trajectories, their planarity, (quasi-)periodicity and 
quantization. The quasi-periodicity of the bounded motions 
is connected in Section 4 to the concept of ``local'' symmetry 
of a spectral problem as first 
introduced by Coulson's school in connection with the Kepler 
problem [24]. The planarity of 
the bounded motions is studied with the help of some new 
formulas, relegated to the appendix, for the torsion (and the 
curvature) of the trajectories of a given hamiltonian system. 
\bigskip
\smallskip

\centerline {\bf 2. The Oscillatory Ring-Shaped Potential}
\bigskip

\noindent {\it 2.1. Generalities}
\smallskip

We shall deal in this section with the potential (energy)
$$
U_Q = 
{1\over 2} \, \Omega^2 \, ({x_1}^2 + {x_2}^2 + {x_3}^2) + 
{1\over 2} \, Q \, {1 \over {{x_1}^2 + {x_2}^2}} 
\qquad  \Omega > 0 \qquad Q > 0 
\eqno (2.1)
$$
which clearly exhibits an $O(2)$ cylindrical symmetry. The limiting case $Q=0$
corresponds to an isotropic harmonic oscillator (in $\grt$) and 
will serve for
testing the results to be obtained.

The potential $U_Q$ is a special case of the potential
$$
V_3 =
\alpha \, ({x_1}^2 + {x_2}^2 + {x_3}^2) + 
\beta \, {1 \over {x_3}^2} +
h({x_2 \over x_1}) \, {1 \over {{x_1}^2 + {x_2}^2}} 
\eqno (2.2)
$$
of Makarov {\it et al}.~[22]. The potential $V_3$ allows the separation of
variables for the Hamilton-Jacobi equation and, also, the 
Schr\"odinger equation in four
systems of coordinates, viz, the spherical, circular cylindrical, prolate
spheroidal and oblate spheroidal coordinates. 
According to Makarov {\it et al}.~[22], there are four 
functionally independent
integrals of motion for $V_3$ and, thus, for $U_Q$. 
More precisely, the four integrals for $U_Q$ can be constructed 
in terms of the relevant Hamilton function and the $A_k$'s listed 
below [2,20,22]. 

(i) Spherical coordinates~:
$$
x_1 = r \sin \theta \cos \varphi \qquad 
x_2 = r \sin \theta \sin \varphi \qquad 
x_3 = r \cos \theta
$$
\rightline{(2.3a)}
$$
A_1 = {\ell_1}^2 + {\ell_2}^2 + {\ell_3}^2 + Q \, {1 \over {\sin^2 \theta}}
\qquad
A_3 = {\ell}_3.
$$

(ii) Circular cylindrical coordinates~:
$$
x_1 = \rho \cos \varphi \qquad 
x_2 = \rho \sin \varphi \qquad 
x_3 = z
$$
\rightline{(2.3b)}
$$
A_2 = {1 \over 2} \, ({p_3}^2 + \Omega^2 {x_3}^2) \qquad
A_3 = {\ell}_3.
$$

(iii) Prolate spheroidal coordinates~:
$$
x_1 = a \sinh \eta \sin \alpha \cos \varphi \qquad
x_2 = a \sinh \eta \sin \alpha \sin \varphi \qquad
x_3 = a \cosh \eta \cos \alpha
$$
\rightline{(2.3c)}
$$
A_4 = {\ell_1}^2 + {\ell_2}^2 + {\ell_3}^2 - a^2 ({p_1}^2 + {p_2}^2)
- a^4 \Omega^2 \sinh^2 \eta \sin^2 \alpha 
+ Q \, {{\sinh^2 \eta - \sin^2 \alpha} \over 
{\sinh^2 \eta \sin^2 \alpha}} \qquad 
A_3 = \ell_3.
$$

(iv) Oblate spheroidal coordinates~:
$$
x_1 = a \cosh \eta \sin \alpha \cos \varphi \qquad
x_2 = a \cosh \eta \sin \alpha \sin \varphi \qquad
x_3 = a \sinh \eta \cos \alpha 
$$
\rightline{(2.3d)}
$$
A_5 = {\ell_1}^2 + {\ell_2}^2 + {\ell_3}^2 + a^2 ({p_1}^2 + {p_2}^2)
+ a^4 \Omega^2 \cosh^2 \eta \sin^2 \alpha 
+ Q \, {{\cosh^2 \eta + \sin^2 \alpha} \over 
{\cosh^2 \eta \sin^2 \alpha}} \qquad 
A_3 = \ell_3.
$$
(It is important to note that, in order to adhere 
to standard notations, terms of type
($p_k$) and ($-M_k$ or $L_k$) in the notation of Makarov 
{\it et al}.~[22] or Kibler and Winternitz [7] are replaced 
here by ($ip_k$) and ($-i\ell_k$), respectively. The variables 
$p_k$ and $\ell_k$ in the present paper stand for linear 
and angular momenta, respectively, in atomic units.)

Returning to Cartesian coordinates, we can write the first 
three integrals of motion as
$$
A_1 = {\ell_1}^2 + {\ell_2}^2 + {\ell_3}^2 + 
Q \, { { {x_1}^2 + {x_2}^2 + {x_3}^2 } \over { {x_1}^2 + {x_2}^2 } } 
\qquad
A_2 = {1 \over 2} \, ( {p_3}^2 + \Omega^2 {x_3}^2 ) \qquad
A_3 = x_1p_2 - x_2p_1.
\eqno (2.4)
$$
The remaining two then satisfy
$$
A_4 = A_1 + 2 a^2 (A_2 - H) \qquad \quad A_5 = A_1 - 2 a^2 (A_2 - H)
\eqno (2.5)
$$
with
$$
H = {1 \over 2} \, ({p_1}^2 + {p_2}^2 + {p_3}^2) + U_Q.
\eqno (2.6)
$$
We see that $\{ H, A_1, A_2, A_3\}$ form an integrity basis for 
all integrals of the considered Hamiltonian system. 

A consequence of Eq.~(2.5) is that any potential that 
allows the separation of variables in spherical and cylindrical 
coordinates will also allow the separation in prolate and 
oblate spheroidal coordinates. This holds both for the 
Hamilton-Jacobi and Schr\"odinger equations. 
\smallskip

\noindent {\it 2.2. Equipotential Surfaces}
\smallskip

The equipotentials corresponding 
to given values of $U_Q$, with $U_Q \ge \Omega \, \sqrt Q$, are 
described (in circular cylindrical coordinates) by 
$$
z = \pm \, {1 \over {\Omega}}
        \, {1 \over {\rho  }} 
        \, \sqrt{- \Omega^2 \, \rho^4 + 2 \, U_Q \, \rho^2 - Q} \qquad
\rho = \sqrt{{x_1}^2 + {x_2}^2} \qquad z = x_3
\eqno (2.7)
$$
with the restrictions
$$
{\rho_<}^2 \le \rho^2 \le {\rho_>}^2
\eqno (2.8)
$$
where
$$
{\rho_<}^2 = {{U_Q - \sqrt{{U_Q}^2 - \Omega^2 Q}} \over \Omega^2} \qquad
{\rho_>}^2 = {{U_Q + \sqrt{{U_Q}^2 - \Omega^2 Q}} \over \Omega^2}.
\eqno (2.9)
$$
All the equipotentials corresponding to bounded values
of $U_Q$ with $U_Q > \Omega \, \sqrt{Q}$ are bounded surfaces which
collapse into simple lines (actually circles in the $z = 0$ plane) for $U_Q=
\Omega \, \sqrt{Q}$. 
Of course, each equipotential reduces to a simple 
sphere, of radius ${\sqrt{2 \, U_0}}/{\Omega}$, in the limiting 
situation ($Q = 0$, $U_0 > 0$).
\smallskip

\noindent {\it 2.3. Trajectories}
\smallskip

We now derive the equations of motion for a particle 
with potential energy $U_Q$. We use here the cylindrical coordinates 
$q \equiv (\rho, \varphi, z)$ 
in which the Hamilton function $H$ for a particle of 
(reduced) mass $\mu = 1$ in the potential $U_Q$ reads
$$
H \equiv H (p,q) =
{1 \over 2} 
\left(
{p_\rho}^2 + {1 \over \rho^2} \, {p_\varphi}^2 + {p_z}^2 
\right)
+ {1 \over 2} \, \Omega^2 \, 
(\rho^2 + z^2) 
+ {1 \over 2} \, Q \, {1 \over \rho^2}
\eqno (2.10)
$$
where $p \equiv (p_\rho = {\dot \rho}, \, 
p_\varphi = \rho^2 {\dot \varphi}, \, p_z = {\dot z})$
are the relevant canonically conjugated momenta. The 
corresponding Hamilton-Jacobi equation can be written
$$
{{\partial S} \over {\partial t}}
+ {1 \over 2}
\left [
\left(
{{\partial S} \over {\partial \rho}}
\right)^2
+ {1 \over {\rho^2}}
\left(
{{\partial S} \over {\partial \varphi}}
\right)^2
+ \left(
{{\partial S} \over {\partial z}}
\right)^2
\right ]
+ {1 \over 2} \, \Omega^2 \, (\rho^2 + z^2)
+ {1 \over 2} \, Q \, {1 \over {\rho ^2}} = 0.
\eqno (2.11)
$$
Obviously, $\varphi$ is a cyclic variable so that we put
$$
p_\varphi = {{\partial S} \over {\partial \varphi}} = m \qquad 
m \in \gr
\eqno (2.12)
$$
and, since the system of the particle in $U_Q$ is conservative, we
look for a solution of $(2.11)$ in the form
$$
S = S_1(\rho) + m \, \varphi + S_2(z) - E \, t.
\eqno (2.13)
$$
This leads to the separated equations
$$
{1 \over 2} 
\left (
{{dS_1} \over {d \rho}} 
\right )^2
+ {1 \over 2} \, \Omega^2 \rho^2
+ {1 \over 2} \, (m^2 + Q) \, {1 \over {\rho^2}}
= K \qquad \quad 
{1 \over 2}
\left (
{{dS_2} \over {dz}} 
\right )^2
+ {1 \over 2} \, \Omega ^2 z^2 = - K + E
\eqno (2.14)
$$
which are coupled via a (positive) separation constant $K$. Equations 
(2.14) show that the only possible motions are bounded motions. They occur 
for
$$
E \ge K > 0 \qquad K^2 - \Omega^2 M^2 \ge 0 
\eqno (2.15)
$$
where
$$
M^2 = m^2 + Q.
\eqno (2.16)
$$
Indeed, Eqs.~(2.15) ensure that
$$
- z_0 \le z(t) \le z_0 \qquad \rho_1 \le \rho(t) \le \rho_2
\eqno (2.17)
$$
with
$$
z_0 = {{\sqrt {2(E - K)}} \over {\Omega}}
\qquad
\rho_1 = {{\sqrt {K - {\sqrt {K^2 - \Omega^2 M^2}}}} \over {\Omega}}
\qquad
\rho_2 = {{\sqrt {K + {\sqrt {K^2 - \Omega^2 M^2}}}} \over {\Omega}}.
\eqno (2.18)
$$
Direct integration of (2.14) yields
$$
\eqalign{
S_1 (\rho) & = {1 \over 2} \, \Omega \, 
{\sqrt {\left( \rho^2 - {\rho_1}^2 \right) 
        \left( {\rho_2}^2 - \rho^2 \right)}} \cr
           & + {1 \over 2} \, {K \over \Omega} \, \sin^{-1}
\left[ {{2 \rho^2 - \left( {\rho_1}^2 + {\rho_2}^2 \right)} 
\over {{\rho_2}^2 - {\rho_1}^2}} \right] 
- {1 \over 2} \, |M| \, \sin^{-1}
\left[ {{ \left( {\rho_1}^2 + {\rho_2}^2 \right) \rho^2 
- 2 {\rho_1}^2 {\rho_2}^2}
  \over { \left( {\rho_2}^2 - {\rho_1}^2 \right) \rho^2 }} \right] \cr
}
\eqno (2.19a)
$$
$$
S_2 (z) = {{E - K} \over \Omega} 
\left [
{z \over z_0} \,
{\sqrt {1 - \left( {z \over z_0} \right)^2}}
+ \sin^{-1}
\left (
{z \over z_0}
\right )
\right ].
\eqno (2.19b)
$$
Equation ($2.19a$) is valid for $K^2 - \Omega^2 M^2 > 0$ ; in the special case
$K^2 - \Omega^2 M^2 = 0$, we have
$$
\rho_1 = \rho_2 \equiv \rho_0 =
{{\sqrt K} \over {\Omega}} =
{{|M|} \over {\sqrt {K}}} =
\sqrt {{|M|} \over {\Omega}} 
\quad \Rightarrow \quad 
S_1 (\rho) = {\rm constant}
\eqno (2.20)
$$
and the motion is constrained to $\rho = \rho_0$. 

The action $S$ follows from (2.13) and (2.19). Therefore, the equations 
of motion
$$
{{\partial S} \over {\partial E}} = \beta_1 \qquad
{{\partial S} \over {\partial K}} = \beta_2 \qquad
{{\partial S} \over {\partial m}} = \beta_3 
\eqno (2.21)
$$
can be derived in a straightforward way. In fact, 
in view of the relations
$$
\beta_1 = {{\partial S_2} \over {\partial E}} - t \qquad
\beta_2 = {{\partial S_1} \over {\partial K}} + 
          {{\partial S_2} \over {\partial K}} =
          {{\partial S_1} \over {\partial K}} - 
          {{\partial S_2} \over {\partial E}} \qquad
\beta_3 = {{\partial S_1} \over {\partial m}} + \varphi
\eqno (2.22)
$$
we choose $t'_0 \equiv - \beta_1$, $t_0 \equiv - (\beta_1 + \beta_2)$ and 
$\varphi_0 \equiv \beta_3$ for constants of motion. Thus, we obtain 
(for $K^2 - \Omega^2 M^2 > 0$) 
$$
t - t'_0 = {1 \over \Omega} \, \sin^{-1}
\left( {z \over z_0} \right)
$$
$$
t - t_0 = {1 \over 2 \Omega} \, \sin^{-1}
\left[ {{2 \rho^2 - \left( {\rho_1}^2 + {\rho_2}^2 \right)} 
\over {{\rho_2}^2 - {\rho_1}^2}} \right]
\eqno (2.23)
$$
$$\varphi - \varphi_0 = {1 \over 2} \, {m \over {|M|}} \, \sin^{-1} 
\left[ {{ \left( {\rho_1}^2 + {\rho_2}^2 \right) \rho^2 
- 2 {\rho_1}^2 {\rho_2}^2} 
  \over { \left( {\rho_2}^2 - {\rho_1}^2 \right) \rho^2 }} \right]
$$
from which we can extract the coordinates $\rho(t)$, $\varphi(t)$ and $z(t)$. 
We finally arrive at
$$
\rho(t) = {1 \over \sqrt 2} \,
\sqrt {{\rho_1}^2 + {\rho_2}^2 + ({\rho_2}^2 - {\rho_1}^2) 
\sin [2 \Omega (t - t_0)]}
\eqno (2.24a)
$$
$$
\varphi(t) = \varphi_0 + {1 \over 2} \, {m \over {|M|}} \, 
\sin^{-1}
\left \{
{{{\rho_2}^2 - {\rho_1}^2 + ({\rho_1}^2 + {\rho_2}^2) 
\sin [2\Omega (t - t_0)]}
\over
 {{\rho_1}^2 + {\rho_2}^2 + ({\rho_2}^2 - {\rho_1}^2) 
\sin [2\Omega (t - t_0)]}}
\right \}
\eqno (2.24b)
$$
$$
z(t) = z_0 \sin [\Omega (t - t'_0)]
\eqno (2.24c)
$$
in agreement with a previous derivation based on functionally 
independent constants of motion [20]. 
Equations ($2.24a$) and ($2.24b$) 
                        are valid for $K^2 - \Omega^2 M^2 > 0$ 
        and also for                  $K^2 - \Omega^2 M^2 = 0$ 
under the condition $\rho_1 = \rho_2 = \rho_0$. Note that in 
($2.24b$), the parameter $m/|M|$ (which is $\pm 1$ in the 
limiting case $Q = 0$) is given by
$$
{m^2 \over M^2} = 
1 - {Q \over {\Omega}^2} \, {1 \over {{\rho_1}^2 {\rho_2}^2}}
\eqno (2.25)
$$
so that the variables $\rho$, $\varphi$ and $z$ depend on the 
six constants $\rho_1$, $\rho_2$, $\varphi_0$, $z_0$, $t_0$ 
and $t'_0$.

At this stage, it is worth mentioning that ($2.24b$) is equivalent to 
$$
\varphi (t) = \varphi_0 - 
{1 \over 2} \, {m \over {|M|}} \, \alpha +
{m \over {|M|}} \tan^{-1} 
\left \{
{{({\rho_1}^2 + {\rho_2}^2) \tan [\Omega (t - t_0)] + {\rho_2}^2 - {\rho_1}^2}
\over
{2 \rho_1 \rho_2}}
\right \}
\eqno (2.26)
$$
with
$$
\sin \alpha = {{{\rho_2}^2 - {\rho_1}^2} \over
               {{\rho_1}^2 + {\rho_2}^2}}
\qquad
\cos \alpha = {{2 \rho_1 \rho_2} \over
{{\rho_1}^2 + {\rho_2}^2}}.
\eqno (2.27)
$$
Indeed, Eq.~(2.26) naturally arises when solving the Newton equations
for the dynamical system under consideration.

As a r\'esum\'e, all the trajectories are bounded. The most general motion
corresponds to
$$
\Omega > 0 \quad \qquad Q > 0 \quad \qquad E > K > \Omega \, |M| > 0.
\eqno (2.28)
$$
It follows from (2.17) that the classical 
trajectories lay between two cylinders of height $2 z_0$ and 
radii 
$\rho_1$ and $\rho_2$. The real constant $m$ is 
the $z$-component $\ell_3$ of the 
angular momentum for the system being studied~; we thus recover 
that $M^2 = {A_3}^2 + Q$ is a constant of motion 
(cf.~Ref.~[22]). Furthemore, 
the positive constant $K$ is another constant of 
motion, namely, $K = E - A_2$. In terms of $\rho_1$, $\rho_2$ and 
$z_0$, 
the three constants of motion $m$, $K$ and $E$ are given by
$$
m^2 = \Omega^2 \, {\rho_1}^2 \, {\rho_2}^2 - Q \qquad
K = {1 \over 2} \, \Omega^2 \, ({\rho_1}^2 + {\rho_2}^2) \qquad
E = {1 \over 2} \, \Omega^2 \, ({\rho_1}^2 + {\rho_2}^2 + {z_0}^2). 
\eqno (2.29)
$$
Finally, it is to be emphasized that the trajectories are neither 
periodic nor planar in general.
\smallskip

\noindent {\it 2.4. Planarity of Trajectories}
\smallskip

The question of planarity of (bounded and nonbounded) 
motions can be tackled by looking at the torsion $\tau$ of the 
trajectories. A planar motion corresponds to $\tau = 0$ along 
the trajectory. For the 
$U_Q$ system, by using the formula for $\tau = - {\rm NUM}/{\rm DEN}$ derived 
in the appendix, we get
$$
\eqalign{
\rho^8 \times {\rm NUM} & = Q \, m \left[ (Q - \Omega^2 \rho^4) {\dot z} 
+ 4 \, \Omega^2 {\rho}^2 (x {\dot x} + y {\dot y}) z \right] \cr
\rho^8 \times {\rm DEN} & = (Q - \Omega^2 \rho^4)^2 (m^2 + \rho^2 {\dot z}^2) 
\cr                     & + 2 \, \Omega^2 \rho^4 (Q - \Omega^2 \rho^4) 
(x {\dot x} + y {\dot y}) z {\dot z} 
+ \left( \Omega^2 \rho^4 \right)^2 ({\dot x}^2 + {\dot y}^2) z^2. \cr
}
\eqno (2.30)
$$
The condition $\tau = 0$ can be seen to be equivalent to
$$
  Q \, m \left[ (Q - \Omega^2 \rho^4) {\dot z} 
+ 4 \, \Omega^2 {\rho}^3 {\dot \rho} z \right] = 
  Q \, m \left[ (Q - \Omega^2 \rho^4) {d \over dt}(z) 
- z {d \over dt}(Q - \Omega^2 \rho^4) \right] = 0 
\eqno (2.31)
$$
with
$$
\rho {\dot \rho} = x {\dot x} + y {\dot y} = \pm 
\sqrt{ \left[ 2 E - {\dot z}^2 - \Omega^2 (\rho^2+z^2) \right] \rho^2 - M^2}.
\eqno (2.32)
$$
Solutions of Eq.~(2.31) are~:

(i) $Q = 0$. In this case the studied system reduces to the 
harmonic oscillator for which of course all trajectories are 
planar (ellipses with the attractive ``sun'' in the center). 

(ii) $z = {\dot z} = 0$. We then have $z_0=0$ in (2.24c) and 
hence $E = K$, see (2.29).

(iii) $m = 0$. The motion is restricted to the plane $\varphi = 
\varphi_0$, see (2.24b) and (2.26), and the limiting radii 
satisfy $\rho_1 \rho_2 = \sqrt{Q}/\Omega$, see (2.29). 

(iv) $\rho^4 = Q/{{\Omega}^2}$. Indeed, the generic solution of 
(2.31) is $Q - \Omega^2 \rho^4 = c_0 z$ with 
$c_0 = {\rm constant}$. However, since $z$ and $\rho$ have 
different periods, see (2.24), the latter solution can only 
hold for $\rho = \rho_1 = \rho_2 = (Q/\Omega^2)^{1/4}$ which 
implies $m=0$, see (2.29), 
and hence either $z_0=0$ (no motion) or $c_0=0$. 

To sum up, excluding the well known harmonic oscillator
corresponding to the limiting case $Q=0$, we find that 
the trajectories are planar only if we have $m=0$ or $E=K$.
\smallskip

\noindent {\it 2.5. Periodicity and Quasi-Periodicity}
\smallskip

From Eq.~(2.23), we easily see that the projections 
onto the $xy$ plane of the trajectories are described by
$$
\rho = {\sqrt 2} \, 
{{\rho_1 \rho_2} \over
\sqrt {{\rho_1}^2 + {\rho_2}^2 - ({\rho_2}^2 - {\rho_1}^2) 
\sin \left[ 2 \, {{|M|} \over m} \, (\varphi - \varphi_0) \right] }}
\eqno (2.33)
$$
and, thus, these projections are not closed in general. Equation (2.33)
shows that the projection onto the $xy$ plane of a given trajectory is 
closed if the condition
$$
{{|M|} \over m} = 
{{k_1} \over {k_2}} \qquad \quad
k_1 \in \grz        \qquad
k_2 \in \grz
\eqno (2.34)
$$
(with ${k_1}^2 > {k_2}^2$) is fulfilled. The situation 
      ${k_1}^2 = {k_2}^2 = 1$
corresponds to the limiting 
case $Q = 0$ for which we know that all trajectories (actually ellipses) 
are planar and periodic. In the situation where (2.34) is satisfied, 
then not only the projection, onto the $xy$ plane, of the 
trajectory is 
periodic, but the trajectory itself is periodic too. The period of
the motion is then
$$
T = k_1 \, T_O \qquad T_O = {2 \pi \over \Omega} 
\eqno (2.35)
$$
where $T_O$ is the oscillator period corresponding to the limiting 
case $Q = 0$. (See also Ref.~[20] for an alternative derivation of (2.35).)

In the general case $Q > 0$, when the condition (2.34) is 
satisfied, we have
$$
m^2 = Q \, {{k_2}^2 \over {{k_1}^2 - {k_2}^2}} \qquad 
{\rho_1}^2 {\rho_2}^2 = {Q \over \Omega^2} \, {{k_1}^2 \over 
{{k_1}^2 - {k_2}^2}} 
\eqno (2.36)
$$
for the periodic motions. 
The requirement that the trajectories be periodic thus leads to 
a ``quantization condition'' for the 
component $\ell_3$ of the angular momentum and the mean geometric 
radius $\sqrt {\rho_1 \rho_2}$. Motions for which (2.34) 
is not satisfied are quasi-periodic motions rather than periodic 
ones. 

The potential energy
$$
{\overline {U_Q}} = {1 \over T_O} \, \int_0^{T_O} \, 
\left [
{1 \over 2} \, \Omega^2 \, (\rho^2 + z^2) + 
{1 \over 2} \, Q \, {1 \over \rho^2}
\right ] \, dt
\eqno (2.37)
$$
averaged over a duration $T_O$ (a common period for $\rho$ and $z$)
can be easily calculated from ($2.24a$) and ($2.24c$). We get
$$
{\overline {U_Q}} = {1 \over 4} \, \Omega^2 \, 
({\rho_1}^2 + {\rho_2}^2 + {z_0}^2) 
+ {1 \over 2} \, Q \, {1 \over \rho_1 \rho_2} 
\quad \Rightarrow \quad 
{\overline {U_Q}} = {1 \over 2} \, E 
+ {1 \over 2} \, Q \, {1 \over \rho_1 \rho_2}
\eqno (2.38)
$$
for periodic and quasi-periodic motions. Thus, the 
virial theorem (${\overline T} = {\overline V} = E/2$) for 
the three-dimensional isotropic harmonic oscillator applies to the $U_Q$ 
system only when $Q=0$.
\smallskip

\noindent {\it 2.6. Particular Cases}
\smallskip

{\it The case $Q = 0$.} As a general check of the correctness of 
($2.24a$), ($2.24b$) (or (2.26)) and ($2.24c$), we can verify, by calculating 
$x = \rho \cos \varphi$ and $y = \rho \sin \varphi$, that the
limiting case $Q = 0$ effectively corresponds to a 
three-dimensional isotropic harmonic oscillator with angular frequency 
$\Omega$. As a further check, it can be verified that for
$Q = 0$ the components of the angular momentum for the considered 
particle are
$$
\eqalign{
\ell_2 & = - {\Omega \over \sqrt {2}} \, z_0 
\sqrt{{\rho_1}^2 + {\rho_2}^2} 
\left \{
\cos (\varphi_0 + {\alpha \over 2}) \cos [\Omega (t'_0 - t_0)] -
\sin (\varphi_0 - {\alpha \over 2}) \sin [\Omega (t'_0 - t_0)]
\right \} \cr
\ell_3 & = m \cr
\ell_1 & = + {\Omega \over \sqrt {2}} \, z_0 
\sqrt{{\rho_1}^2 + {\rho_2}^2} 
\left \{ 
\sin (\varphi_0 + {\alpha \over 2}) \cos [\Omega (t'_0 - t_0)] +
\cos (\varphi_0 - {\alpha \over 2}) \sin [\Omega (t'_0 - t_0)]
\right \} \cr
}
\eqno (2.39)
$$
indicating that the elliptic trajectory corresponding to 
$\rho_1$, $\rho_2$, $\varphi_0$, $z_0$, $t_0$ and $t'_0$ 
is in the plane perpendicular 
to the constant vector $(\ell_1, \ell_2, \ell_3)$.

{\it The case $K = E$.} Equations (2.18) and ($2.24c$) show that the 
trajectories are in
the $xy$ plane when $E = K$, a situation that corresponds to 
case (ii) of section 2.4. They are described by (2.33) and thus are not 
closed in general. However, the planar trajectories corresponding to $E = K$
become closed if the condition (2.34) is satisfied.

{\it The case $m = 0$.} Case (iii) of section 2.4 
indicates that the trajectories are planar for $m=0$ and, from 
Eqs.~(2.24), we see that they are in the $\rho z$ plane 
corresponding to $\varphi(t) = \varphi_0$. All the trajectories 
($\rho(t)$, $z(t)$) are periodic, of period 
$T_O$.

{\it The case $K = \Omega |M|$.} The 
two cylinders of radii $\rho_1$ and
$\rho_2$ collapse into a single one, of radius $\rho_0$, when 
$K = \Omega |M|$ ; therefore, the trajectories (neither closed 
nor planar in general) are
on a cylinder of height $2 z_0$ and radius 
$\rho_0$ (see (2.17), (2.18) and (2.20)). Then, Eqs.~(2.24) 
lead to the Cartesian coordinates 
$$
x(t)=\rho_0 \, \cos \left( \Omega \, {{m} \over {|M|}} \, t+\varphi'_0 \right) 
\quad
y(t)=\rho_0 \, \sin \left( \Omega \, {{m} \over {|M|}} \, t+\varphi'_0 \right)
\quad
z(t)=z_0 \, \sin [\Omega (t - t'_0)]
\eqno (2.40)
$$
where $\varphi'_0$ is some new constant. The motion thus results from the
combination of a circular motion of angular frequency
$$
\omega = \Omega \, { {m} \over {|M|} } = { {m} \over {\rho_0}^2 }
\eqno (2.41)
$$
in the $xy$ plane and of an oscillatory motion of angular frequency $\Omega$ in
the $z$ direction. (Without loss of generality, we can take $m > 0$ ; the 
case $m = 0$ is trivial.) In the limiting case $Q = 0$, the 
trajectory reduces to an 
ellipse, the projection of which on the $xy$ plane is a circle of
radius $\rho_0$.

In the case $K = \Omega |M|$ and $Q > 0$, the trajectories are not closed 
in general. However, if we further assume that the condition 
(2.34) is satisfied, the 
trajectories become closed and of period $T = k_1 T_O$. They 
are given by (2.40) and (2.41) with
$$
{\Omega \over \omega} = {k_1 \over k_2} \qquad \quad 
\rho_0 =
\root\scriptstyle 4\of { {Q} \over {{\Omega}^2} } \, 
\root\scriptstyle 4\of { {{k_1}^2} \over {{k_1}^2 - {k_2}^2} }
\eqno (2.42)
$$
with $k_1 > k_2$, $k_1 \in \grn - \{0,1\}$ and $k_2 \in \grn - \{0\}$.
\smallskip

\noindent {\it 2.7. Semi-Classical Quantization}
\smallskip

We now use the Bohr-Sommerfeld-Kramers quantization conditions 
to derive, in a semi-classical way, the quantum mechanical spectrum 
for the $U_Q$ system. Here, these conditions read 
$$
\oint p_{\rho} d\rho = (n_{\rho} + \epsilon_{\rho}) 2 \pi \quad 
n_{\rho} \in \grn \qquad \quad 
\oint p_{z}    dz    = (n_{z}    + \epsilon_{z})    2 \pi \quad
n_{z}    \in \grn
\eqno (2.43)
$$
where ${\epsilon}_{\rho}$
and   ${\epsilon}_{z}   $ stand for (rational) numbers to 
reconcile the old and new theories of quanta. (In the case of a 
one-dimensional harmonic oscillator, we know that ${\epsilon}_{z} = 1/2$.) 
In other words, we have 
$$
2 \left [ S_1 (\rho_2) - S_1 (\rho_1) \right ] =
(n_{\rho} + \epsilon_{\rho}) 2 \pi \qquad \quad
2 \left [ S_2 (z_0) - S_2 (-z_0) \right ] =
(n_{z}    + \epsilon_{z})    2 \pi 
\eqno (2.44)
$$
and, from (2.19), we obtain
$$
{1 \over 2} {K \over \Omega} - {1 \over 2} |M| = 
n_{\rho} + \epsilon_{\rho} \qquad \quad
{{E - K} \over {\Omega}} = 
n_{z} + \epsilon_{z}. 
\eqno (2.45)
$$
Consequently, the eigenvalue $E$ is given by 
$$
E = \left( |M| + 2 n_{\rho} + n_{z} + {3 \over 2} \right) \Omega \qquad 
|M| = \sqrt{m^2 + Q} \qquad m \in \grz \qquad
                       n_{\rho} \in \grn \qquad
                          n_{z} \in \grn
\eqno (2.46)
$$
in agreement with the result by Quesne [2]. In fact, 
Eq.~(2.46) demands that 
${\epsilon}_z + 2 {\epsilon}_{\rho} = 3/2$, a relation 
which is reminiscent of the three-dimensional isotropic harmonic 
oscillator.
\smallskip
\bigskip

\centerline {\bf 3. The Coulombic Ring-Shaped Potential}
\bigskip

\noindent {\it 3.1. Generalities}
\smallskip

The coulombic ring-shaped, or Hartmann, potential (energy) is
$$
V_Q = - Z \, {1 \over \sqrt{{x_1}^2 + {x_2}^2 + {x_3}^2}} + {1 \over 2} \,
Q \, {1 \over {{x_1}^2 + {x_2}^2}} \qquad Z > 0 \qquad Q > 0 
\eqno (3.1)
$$
where $Z = \eta \, \sigma^2$ and $Q = q \, \eta^2 \, \sigma^2$ in the
notation of Hartmann [1] and of Kibler and N\'egadi [3]. Such 
an $O(2)$ invariant potential reduces to an attractive Coulomb
potential (in $\gr^3$) in the limiting case $Q = 0$ and this will 
prove useful for checking purposes. 

Clearly, $V_Q$ is a special case of the potential (in spherical 
coordinates) 
$$
V_4 = \alpha \, {1 \over r} + 
\beta \, {{\cos \theta} \over {r^2 \sin^2 \theta}} + 
h (\tan \varphi) \, {1 \over {r^2 \sin^2 \theta}}
\eqno (3.2)
$$
introduced by Makarov {\it et al}.~[22]. Therefore, the 
Schr\"odinger equation and, thus, the
Hamilton-Jacobi equation for $V_Q$ are separable in spherical 
and parabolic rotational coordinates. In this
respect, let us simply recall that there are four functionally independent
integrals of motion for $V_Q$ which can be obtained from the relevant
Hamilton function and the $B_k$'s given below [7,20,22].

(i) Spherical coordinates~:
$$
x_1 = r \sin \theta \cos \varphi \qquad 
x_2 = r \sin \theta \sin \varphi \qquad 
x_3 = r \cos \theta
$$
\rightline{(3.3)}
$$
B_1 = {\ell_1}^2 + {\ell_2}^2 + {\ell_3}^2 + Q \, {1 \over {\sin^2 \theta}}
\qquad
B_2 = {\ell}_3.
$$

(ii) Parabolic rotational coordinates~:
$$
x_1 = \sqrt {ab}  \, \cos \varphi \qquad
x_2 = \sqrt {ab}  \, \sin \varphi \qquad
x_3 = {1 \over 2} \, (a - b)
$$
\rightline{(3.4)}
$$
B_3 = {1 \over 2} \left( 
\ell_1 p_2 + p_2 \ell_1 - \ell_2 p_1 - p_1 \ell_2 
+ 2 \, Z \, {{a - b} \over {a + b}} - Q \, {{a - b} \over {ab}} 
\right) \qquad B_2 = \ell_3.
$$
(The constant $B_3$ is given in a quantum mechanical 
form. The factor $+2[\ldots]$ in the corresponding expression 
of Ref.~[22] for $B_3$ should be changed into 
$-2[\ldots]$.)
\smallskip

\noindent {\it 3.2. Equipotential Surfaces}
\smallskip

The equipotentials are formally given (in 
circular cylindrical coordinates) by 
$$
z = \pm \, {\rho \over {Q - 2 V_Q \rho^2}} \, 
\sqrt {4 Z^2 \rho^2 - (Q - 2 V_Q \rho^2)^2} \qquad
\rho = \sqrt{{x_1}^2 + {x_2}^2} \qquad z = x_3.
\eqno (3.5)
$$
In contradistinction with $U_Q$, three cases 
must be considered for the potential $V_Q$.

(i) The case $-(1/2) (Z^2/Q) < V_Q < 0$~: The equipotentials are given
by (3.5) with the restrictions
$$
\rho_< \, \le \, \rho \, \le \, \rho_> \qquad \quad 
\rho_< = {{Z - \sqrt {Z^2 + 2 Q V_Q}} \over {- 2 V_Q}} \qquad \quad 
\rho_> = {{Z + \sqrt {Z^2 + 2 Q V_Q}} \over {- 2 V_Q}} 
\eqno (3.6)
$$
and each equipotential is a bounded surface which reduces to 
a circle in the $z=0$ plane for $V_Q = - (1/2) (Z^2/Q)$. In the 
limiting case $Q = 0$, each equipotential for $V_0 < 0$ becomes 
a sphere of radius $- Z /V_0$. 

(ii) The case $V_Q = 0$~: The equipotentials are obtained from 
$$
z = \pm \, {1 \over Q} \, \rho \, \sqrt {4 Z^2 \rho^2 - Q^2} 
\eqno (3.7)
$$
with the restriction
$$
\rho \ge {1 \over 2} \, {{Q} \over {Z}}
\eqno (3.8)
$$
and thus the equipotentials are not bounded surfaces in this case. (The
limiting case $Q = 0$ yields of course a sphere of infinite radius.)

(iii) The case $V_Q > 0$~: The equipotentials are given by (3.5) with 
the restriction
$$
\rho_< \, \le \, \rho \, < \, \rho_> \qquad \quad 
\rho_< = {{ -Z +  \sqrt {Z^2 + 2 Q V_Q}} \over {2 V_Q}} \qquad \quad 
\rho_> = {1 \over \sqrt{2}} \sqrt {Q \over V_Q}
\eqno (3.9)
$$
and here again all the equipotentials are bounded surfaces.
\smallskip

\noindent {\it 3.3. Trajectories}
\smallskip

3.3.1. Preliminaries. We shall solve the 
Hamilton-Jacobi equation for  $V_Q$  in spherical coordinates. The 
general pattern to be followed for $V_Q$ resembles 
the one for $U_Q$ so that we shall mention only the basic 
steps.

The Hamilton-Jacobi equation for a particle of reduced mass $\mu = 1$ in 
$V_Q$ can be written as
$$
{{\partial S} \over {\partial t}} + 
{1 \over 2} \,
\left [
\left (
{{\partial S} \over {\partial r}}
\right )^2 +
{1 \over {r^2}} \,
\left (
{{\partial S} \over {\partial \theta}}
\right )^2 +
{1 \over {r^2 \sin^2 \theta}} \,
\left (
{{\partial S} \over {\partial \varphi}}
\right )^2
\right ] -
Z \, {1 \over r} + {1 \over 2} \,
Q \,
{1 \over {r^2 \sin^2 \theta}} = 0
\eqno (3.10)
$$
and a solution of the type
$$
S = S_1 (r) + S_2 (\theta) + m \, \varphi - E \, t
\eqno (3.11)
$$
is easily found by solving the two equations
$$
r^2
\left (
{{dS_1} \over {dr}}
\right )^2 - 2 E r^2 - 2 Z r = - K
\qquad \quad 
\left (
{{dS_2} \over {d \theta}}
\right )^2
+ (m^2 + Q) {1 \over {\sin^2 \theta}} = K
\eqno (3.12)
$$
where $K$ is a (positive) separation constant. Equations (3.12) show that
two kinds of motions may occur here~: nonbounded 
motions for $E \ge 0$ and bounded motions for
$$
- {1 \over 2} \, {{Z^2} \over {K}} \le E < 0.
\eqno (3.13)
$$
We shall restrict ourselves to the bounded motions and to the 
separatrix which corresponds to $E = 0$. As in the case for $U_Q$, 
the equations of motion are given by (2.21) and we 
shall use the constant $M$ defined via (2.16).
\smallskip

3.3.2. Bounded motions. They are confined in a region delimited by
$$
r_1 \le r(t) \le r_2 \qquad
r_1 = {{Z - \sqrt {Z^2 + 2 E K}} \over {- 2 E}} \qquad
r_2 = {{Z + \sqrt {Z^2 + 2 E K}} \over {- 2 E}} 
$$
\rightline{(3.14)}
$$
\theta_0 \le \theta(t) \le \pi - \theta_0 \qquad 
\quad \sin \theta_0 = { |M| \over {\sqrt{K}} }.
$$
By assuming that $- (1/2) (Z^2 / K) < E < 0$, Eqs.~(3.12) 
admit the solutions
$$
\eqalign{
S_1 (r) & = (- 2E)^{1/2} \sqrt{(r - r_1) (r_2 - r)} \cr 
        & + (- 2E)^{-1/2} \, Z \, \sin^{-1} 
\left[ {{2 r - (r_1 + r_2)} \over {r_2 - r_1}} \right] 
             - \sqrt{K} \sin^{-1} \left[ 
{{(r_1 + r_2) r - 2 r_1 r_2} \over {(r_2 - r_1) r}} \right] \cr
}
\eqno (3.15a)
$$
$$
\eqalign{
S_2 (\theta) = & {|M| \over 2} \left\{ 
  \sin^{-1} \left[ {1 \over \cos \theta_0} \left( -1 + 
{\sin^2 \theta_0 \over {1 - \cos \theta}} \right) \right]
- \sin^{-1} \left[ {1 \over \cos \theta_0} \left( -1 + 
{\sin^2 \theta_0 \over {1 + \cos \theta}} \right) \right] \right\} \cr 
               & - \sqrt{K} \, 
\sin^{-1} \left( {{\cos \theta} \over {\cos \theta_0}} \right). \cr 
}
\eqno (3.15b) 
$$
Putting $t_0 \equiv - \beta_1$, $\beta_0 \equiv - 2 \sqrt{K} 
\beta_2$ and $\varphi_0 \equiv \beta_3$, we get from (2.21) 
$$
t - t_0  = - (- 2E)^{-1/2} \sqrt{(r - r_1) (r_2 - r)} 
+ Z (- 2E)^{-3/2} \sin^{-1} 
\left[ {{2 r - (r_1 + r_2)} \over {r_2 - r_1}} \right] 
\eqno (3.16a)
$$
$$
\beta_0 = \sin^{-1} \left( {{\cos \theta} \over {\cos \theta_0}} \right) + 
\sin^{-1} \left[ 
{{(r_1 + r_2) r - 2 r_1 r_2} \over {(r_2 - r_1) r}} \right]
\eqno (3.16b)
$$
$$
\varphi - \varphi_0 = {1 \over 2} {m \over \vert M \vert} \left\{ 
  \sin^{-1} \left[ {1 \over \cos \theta_0} \left( -1 + 
{\sin^2 \theta_0 \over {1 + \cos \theta}} \right) \right]
- \sin^{-1} \left[ {1 \over \cos \theta_0} \left( -1 + 
{\sin^2 \theta_0 \over {1 - \cos \theta}} \right) \right] 
\right\}.
\eqno (3.16c)
$$
Equation ($3.16b$) is amenable to the form
$$
r \cos \theta = {{\cos {\theta_0}} \over {r_2 - r_1}}
\left\{ [2 r_1 r_2 - (r_1 + r_2) r] \cos \beta_0 + 2 \sqrt{r_1 r_2} 
\sqrt{(r - r_1) (r_2 - r)} \sin \beta_0 \right\}
\eqno (3.17)
$$
and, therefore, Eqs.~(3.16) are in accordance with the 
ones derived by Kibler and Winternitz [20] from functionally 
independent constants of motion. Equations (3.16) are valid for 
$- (1/2) (Z^2 / K) < E < 0$. The case $E = - (1/2) (Z^2 / K) < 0$ 
deserves a particular study (see section 3.6). In Eq.~($3.16c$), 
note that the parameter $m/|M|$ (which is $\pm 1$ in 
the limiting case $Q = 0$) can be obtained from
$$
{m^2 \over M^2} = 1 - {1 \over 2} \, {Q \over Z} \, 
{{r_1 + r_2} \over {r_1 r_2}} {1 \over {\sin^2 \theta_0}}
\eqno (3.18)
$$
so that the variables $r$, $\theta$ and $\varphi$ depend on 
the six constants $r_1$, $r_2$, $\theta_0$, $\varphi_0$, $t_0$ 
and $\beta_0$.

To sum up, the most general bounded motion corresponds to
$$
Z> 0 \quad \qquad Q > 0 \quad \qquad - {1 \over 2} \, {Z^2 \over K} < E < 0
\eqno (3.19)
$$
and all finite trajectories take place 
between two spheres of radii $r_1$ and $r_2$ (see (3.14)). The 
interpretation of the constant $m$ (or $M$) is
similar to the one for $U_Q$~: we have $M^2 = {B_2}^2 + Q$. 
Furthermore, it can be verified that $K = B_1$. In terms of  
$r_1$, $r_2$ and $\theta_0$, the three constants of motion 
$m$, $K$ and $E$ can be deduced from
$$
m^2 = 2 \, Z \, {{r_1 r_2} \over {r_1 + r_2}} \sin^2 {\theta_0} 
- Q \qquad
\quad K = 2 \, Z \, {{r_1 r_2} \over {r_1 + r_2}} \qquad
\quad E = - Z \, {1 \over {r_1 + r_2}}.
\eqno (3.20)
$$
Here again, the finite trajectories are 
neither periodic nor planar in general.
\smallskip

3.3.3. The separatrix. By taking $E=0$, Eqs.~(3.12) lead to
$$
r(t) \ge r_{\rm min} \qquad r_{\rm min} = {1 \over 2} {K \over Z} \qquad 
\theta_0 \le \theta(t) \le \pi - \theta_0 \qquad 
\sin \theta_0 = { |M| \over {\sqrt{K}} }
\eqno (3.21)
$$ 
so that the motions are nonbounded. From (3.11) and (3.12), it 
appears that the coordinate $\varphi(t)$ is still given by ($3.16c$) 
when $E = 0$. In addition, the coordinate $r(t)$ can be determined 
from the cubic equation
$$
r^3 + 3 \, r^2 \, r_{\rm min} - 4 \, {r_{\rm min}}^3 - 
{9 \over 2} \, Z \, (t-t_0)^2 = 0
\eqno (3.22)
$$
while the coordinate $\theta(t)$ follows from
$$
\beta_0 = \sin^{-1} \left ( {{\cos \theta} \over {\cos \theta_0}}
\right ) + 2 \, \tan^{-1} \left ( \sqrt {{r \over r_{\rm min}} - 1} 
\right )
\eqno (3.23)
$$
or alternatively
$$
r \, \cos \theta = \cos \theta_0 
\left [ \left ( 2 \, r_{\rm min} - r \right ) \sin \beta_0 - 2 \, 
\sqrt{r_{\rm min}} \, \sqrt {r - r_{\rm min}} \, \cos \beta_0 \right ] 
\eqno (3.24)
$$
which parallels Eq.~(3.17).
\smallskip

\noindent {\it 3.4. Planarity of Trajectories}
\smallskip

The torsion $\tau$ for the $V_Q$ system can be calculated from 
the general formula $\tau = - {\rm NUM}/{\rm DEN}$ given in the appendix. We 
thus get
$$
\eqalign{
\rho^8 \times {\rm NUM} & = Q \, m \left\{ 
\left( Q - Z {\rho^4 \over r^3} \right) {\dot z} + Z {\rho^2 \over r^5} 
[(x {\dot x} + y {\dot y}) (r^2 + 3 z^2) - 3 z {\dot z} \rho^2] z \right\} \cr
\rho^8 \times {\rm DEN} & = 
\left( Q - Z {\rho^4 \over r^3} \right)^2 (m^2 + \rho^2 {\dot z}^2) \cr
                        & + 2 Z {\rho^4 \over r^3} 
\left( Q - Z {\rho^4 \over r^3} \right) (x {\dot x} + y {\dot y}) z {\dot z}
+ \left( Z {\rho^4 \over r^3} \right)^2 ({\dot x}^2 + {\dot y}^2) z^2. \cr
}
\eqno (3.25)
$$
The condition $\tau = 0$ for planarity amounts to solve
$$
\eqalign{
Q \, m \left[ \left( Q - Z {\rho^4 \over r^3} \right) {\dot z} 
\right. & \left. 
+ Z {\rho^3 \over r^4}
( 4 \, {\dot \rho} r 
- 3 \, {\dot r} \rho ) z \right] \cr
& = Q \, m \left[ \left( Q - Z {\rho^4 \over r^3} \right) {d \over dt}(z)
- z{d \over dt} \left( Q - Z {\rho^4 \over r^3} \right) \right] = 0 
\cr 
} 
\eqno (3.26)
$$
with
$$
\eqalign{
\rho r (4 {\dot \rho} r - 3 {\dot r} \rho)  & = 
(x {\dot x} + y {\dot y}) (r^2 + 3 z^2) - 3 z {\dot z} \rho^2 \cr
\rho {\dot \rho} = x {\dot x} + y {\dot y}  & = \pm 
\sqrt{ \left( 2 E - {\dot z}^2 + 2 Z {1 \over r} \right) \rho^2 - M^2}. \cr
}
\eqno (3.27)
$$
(In (3.25) -- (3.27), $r$ and $\rho$ refer to 
spherical and cylindrical coordinates, respectively.) Equation 
(3.26) is satisfied for all $t$ in each of the following cases~:

(i) $Q = 0$. This is the case of the Kepler planetary system in 
which of course all trajectories are planar, namely, 
hyperbolas, or a parabola, or ellipses with the sun in one 
focus. 

(ii) $z = {\dot z} = 0$. The integrals of motion satisfy 
$|M| = \sqrt{K}$ and $\theta_0 = \pi/2$.

(iii) $m = 0$. The motion is restricted to the plane $\varphi = 
\varphi_0$ and we have $|M| = \sqrt{Q}$. 

(iv) $\rho^4 = (Q/Z) \, r^3$. Indeed, 
case (iv) implies that $ m = 0 $ (at least for 
bounded motions). Furthermore, the generic solution of (3.26) 
requires that $Q - Z (\rho^4/r^3) = c_0 z$ with 
$c_0 = {\rm constant}$, a relation which does not hold for $t$ 
arbitrary (except for the static case corresponding to $z = 0$ 
and $\rho^4 = (Q/Z) r^3$). 

Consequently, in addition to the limiting case $Q=0$ (case 
(i)), the bounded motions are planar only if 
$|M|= \sqrt{K}$ (case (ii)) or $m=0$ (case (iii)).
\smallskip

\noindent {\it 3.5. Periodicity and Quasi-Periodicity}
\smallskip

In the case of finite motions, Eq.~($3.16a$) allows to 
obtain $r$ as function of $t$ by means of a
transcendental equation. The latter equation indicates that $r(t)$ is a
periodic function of period
$$
T_C = 2 \, \pi \, Z \, (- 2E)^{-{3/2}}
\eqno (3.28)
$$
a result in agreement with the fact that the parabolic 
coordinates $a(t)$ and $b(t)$ (with $r = (a + b)/2$) are also of 
period $T_C$ [7]. It should be noted 
that the period $T_C$ bears exactly the same form as the one 
for the Coulomb-Kepler problem that corresponds to the limiting 
case $Q = 0$. Further, from ($3.16b$) or (3.17) it is possible to show 
that $\cos [\theta(t)]$ (and therefore $\sin [\theta (t)]$) is a 
periodic function of period $T_C$. This result 
agrees with our previous work [7] due to the 
passage formulas $\cos \theta = (a - b)/(a + b)$ and 
$\sin \theta = 2 {\sqrt {ab}} / (a + b)$. Finally, the 
consideration of ($3.16c$) leads to
$$
\varphi (t + k_1 T_C) = 
\varphi (t) + 2 \, k_1 \, \pi \, {m \over {|M|}} 
\qquad k_1 \in \grz
\eqno (3.29) 
$$
and, therefore, the (global) motion is periodic of period 
$$
T = k_1 \, T_C
\eqno (3.30)
$$
if the ``quantization'' condition (2.34) for $|M|/m$ is fulfilled 
(see also Ref.~[20]). As a check, in the 
limiting case $Q=0$, the global period $T$ of the motion is 
simply $T_C$.

In the general case $Q > 0$, when the condition (2.34) is 
satisfied, we have the following ``quantized'' expressions
$$
m^2 = Q \, {{k_2}^2 \over {{k_1}^2 - {k_2}^2}} \qquad \quad 
{{r_1 r_2} \over {r_1 + r_2}} \, \sin^2 \theta_0 = {1 \over 2} 
\, {Q \over Z} \, {{k_1}^2 \over {{k_1}^2 - {k_2}^2}}
\eqno (3.31)
$$
for the periodic bounded motions. Here again, the bounded 
motions for which the condition (2.34) is not satisfied are quasi-periodic 
motions. 

The potential energy $V_Q$ averaged over the period $T_C$ (of 
$r$ and $\sin \theta$), namely, 
$$
{\overline {V_Q}} = {1 \over T_C} \, {\int_{0}}^{T_C} \, 
\left ( - Z \, {1 \over r} 
+ {1 \over 2} \, Q \, {1 \over {r^2 \sin^2 \theta}} \right ) \, dt 
\eqno (3.32)
$$
cannot be calculated easily from ($3.16a$) and 
($3.16b$). However, we can calculate in a straightforward way 
the virial of the $V_Q$ system. This leads to 
$$
{\overline {V_Q}} = 2 \, E + {1 \over 2} \, Q \, {1 \over T_C}
\int_0^{T_C} {dt \over {r^2 \sin^2 \theta}} 
                  = 2 \, E +                Q \, {1 \over {r_1 + r_2}} 
\, {1 \over {\sqrt{r_1 r_2}}} \, {1 \over {\sin \theta_0}} 
\eqno (3.33)
$$
for periodic and quasi-periodic bounded motions. Therefore, 
the virial theorem 
(${\overline T} = - {\overline V}/2 = - E$) for the three-dimensional 
Coulomb-Kepler problem applies to the $V_Q$ system only when $Q = 0$.
\smallskip

\noindent {\it 3.6. Particular Cases}
\smallskip

{\it The case $K = M^2$.} Equations (3.12) show that the 
trajectories are in 
the $xy$ plane when $\vert M \vert = \sqrt{K}$, a situation that corresponds 
to case (ii) of section 3.4. The finite trajectories are then described by 
$$
\rho = 2 \, 
{{\rho_1 \rho_2} \over 
{\rho_1 + \rho_2 - (\rho_2 - \rho_1) 
\sin \left[ {|M| \over m} \, (\varphi - \varphi_0) \right]}}
\eqno (3.34)
$$
where $\rho_1$ and $\rho_2$ are given by (3.14) with $r \equiv 
\rho$. These finite trajectories are closed only if the 
condition (2.34) is satisfied. On the other hand, the 
separatrix corresponds to 
$$
\rho = \rho_{\rm min} \left\{ 1 + 
\tan^2 \left[ {1 \over 2} \, {|M| \over m} \, (\varphi - \varphi_0) \right] 
\right\} 
\eqno (3.35)
$$ 
where $\rho_{\rm min}$ is given by (3.21) with $r \equiv 
\rho$.

{\it The case $m = 0$.} The trajectories are planar 
for $m=0$ (cf. case (iii) of section 3.4). They can be obtained in 
the plane $\varphi(t) = \varphi_0$ from ($3.16a$) and (3.17) and 
are all periodic, of period $T_C$.

{\it The case $ 2 E K = - Z^2 $.} The finite 
trajectories are restricted to a sphere of radius
$$
r_1 = r_2 \equiv r_0 = {K \over Z} = {Z \over {-2E}} = \sqrt{K \over {-2E}}
\eqno (3.36)
$$
when $E = - (1/2)(Z^2/K)$. These trajectories (not closed in general) 
are described by
$$
r = r_0 
\quad 
z = r_0 \, \cos \theta_0 \, \cos \left( 2 \pi {t \over T_C} + \psi_0 \right) 
\quad 
\varphi = {m \over |M|} \, 
\tan^{-1} \left[ {{\tan \left( 2 \pi {t \over T_C} + \psi_0 \right)} 
\over {\sin \theta_0}} \right] 
\eqno (3.37)
$$
where $\psi_0$ denotes a new constant. They are periodic 
only if the condition (2.34) for $|M|/m$ is satisfied.
\smallskip

\noindent {\it 3.7. Semi-Classical Quantization}
\smallskip

The Bohr-Sommerfeld-Kramers quantization conditions
$$
\oint p_{r} dr           = (n_{r}      + \epsilon_{r})      2 \pi \quad 
n_{r}      \in \grn \qquad \quad
\oint p_{\theta} d\theta = (n_{\theta} + \epsilon_{\theta}) 2 \pi \quad
n_{\theta} \in \grn
\eqno (3.38)
$$
applied to the $V_Q$ system lead to 
$$
2 \left [ S_1 (r_2)              - S_1 (r_1)        \right ] =
(n_{r}      + \epsilon_{r})      2 \pi \qquad \quad
2 \left [ S_2 (\pi - {\theta}_0) - S_2 ({\theta}_0) \right ] =
(n_{\theta} + \epsilon_{\theta}) 2 \pi. 
\eqno (3.39)
$$
By using Eqs.~(3.15), we get 
$$
  (-2E)^{-1/2} Z - \sqrt {K} = n_{r}      + \epsilon_{r} \qquad \quad
- |M| + \sqrt {K}            = n_{\theta} + \epsilon_{\theta}
\eqno (3.40)
$$
from which we recover the eigenvalue 
$$
E = - {1 \over 2} \, 
{{Z^2} \over {(|M| + n_{r} + n_{\theta} + 1)^2}} \qquad
|M| = \sqrt{m^2 + Q} \qquad m \in \grz \qquad
                          n_{r} \in \grn \qquad
                     n_{\theta} \in \grn
\eqno (3.41)
$$
as first derived by Hartmann [1]. Equation (3.41) 
implies that we must take ${\epsilon}_{\theta} + {\epsilon}_r = 1$, 
a relation which reflects the occurrence of a pair of two-dimensional 
harmonic oscillators in the treatment of the Coulomb-Kepler 
problem by means of the Kustaanheimo-Stiefel transformation [3,7]. 
\smallskip
\bigskip

\centerline {\bf 4. Closing Remarks} 
\bigskip

We have concentrated in this paper on the classical 
motions for two super-integrable systems, viz, the $U_Q$ and 
$V_Q$ systems. The motions are always bounded for $U_Q$ and can 
be bounded or nonbounded for $V_Q$ 
just as in the limiting case $Q=0$ that corresponds to the 
oscillator and the Coulomb-Kepler systems in three dimensions. 
The bounded trajectories are confined inside simple surfaces~: 
cylinders for $U_Q$ and spheres for $V_Q$. 
Both for $U_Q$ and $V_Q$, the trajectories are planar 
only in a few situations~: (i) naturally, when $Q=0$, (ii) when 
a specific relation ($E = K$ for $U_Q$ and $|M| = \sqrt{K}$ for 
$V_Q$) exists between two constants of motion, and (iii) when 
the constant of motion $\ell_3$ vanishes. In addition, 
all bounded trajectories are quasi-periodic rather than 
periodic in general and become truly periodic if a constraint 
condition on $\ell_3$ 
(Eq.~(2.34)), the same for $U_Q$ and $V_Q$, is 
satisfied. In fact, there is an infinity of periodic 
trajectories in the neighborhood of a given quasi-periodic 
trajectory since the set of the rational numbers is dense in 
the set of real numbers. 

From a quantum mechanical point of view, the $U_Q$ system leads 
to a discrete spectrum while the spectrum for the $V_Q$ system
comprises a discrete part, a zero point and a continuum. The 
Hamilton-Jacobi approach developed in sections 2 and 3, as a 
complement to the earlier study of Kibler and Winternitz 
[20], has allowed us to re-derive in a semi-classical way the 
discrete spectra for $U_Q$ and $V_Q$. For both potentials, the 
accidental degeneracies corresponding to fixed values of $|M|$ 
and $E$ are described by a unique dynamical invariance algebra, 
isomorphic to $su(2)$, as shown by Quesne [2] for the $U_Q$ 
system       and by Kibler and Winternitz [7] for the $V_Q$ 
system. 
This result is quite remarkable, especially in view of 
the fact that, in the limiting case $Q=0$, the dynamical 
invariance algebra is isomorphic to $su(3)$ for $U_{Q=0}$ and 
                                 to $so(4)$ for $V_{Q=0}$. 

As was mentioned by Kibler and Winternitz [7], the 
$V_Q$ system may for some energy levels exhibit higher 
degeneracies than those explained by the $SU(2)$ dynamical 
invariance group. These are levels characterized by 
the triplets ($m, n_r, n_{\theta}$)
and ($m', n_r', n_{\theta}'$), with $m' \ne \pm m$, satisfying 
$$
4 Q = {1 \over I^2} [I^2 - (m + m')^2] [I^2 - (m - m')^2] 
\qquad I = n_r' + n_{\theta}' - n_r - n_{\theta} \ne 0. 
\eqno (4.1)
$$
A similar result holds for the $U_Q$ system with 
$I = 2 n_{\rho}' + n_z' - 2 n_{\rho} - n_z$. 

Such degeneracies are ``local'' in that they are restricted to 
part of the energy spectrum of the system (cf.~Ref.~[24]). The quantum 
mechanical operators $X_A$ related to a ``local'' symmetry 
would not commute with the Hamiltonian but would satisfy 
commutation relations of the type 
$$
[X_A,H] \, = \, \lambda (H-E_A)
\eqno (4.2)
$$
where $E_A$ are the energies of the levels for which the accidental 
degeneracy is observed. Since $I$, $m$, and $m'$ are integers, 
Eq.~(4.1) clearly imposes a constraint on the values of the 
coupling constant $Q$ for which additional accidental 
degeneracy can occur. This type of constraint is reminiscent of 
Eqs.~(2.36) and (3.31) that also impose conditions on $Q$. 

As pointed out 
recently by Moshinsky {\it et al}.~[25], the problem of finding a 
group theoretical explanation of accidental degeneracy does not 
have an algorithmic solution. 
It is interesting to speculate about the relation between the 
periodicity condition (2.34) and the local degeneracies and 
we plan to return to this problem in the future. 
\bigskip
\smallskip

\vfill\eject

\centerline {\bf Appendix~: Torsion and Curvature for Hamiltonian Systems}
\bigskip

The purpose of this appendix is to derive closed 
formulas, in Cartesian coordinates, for the torsion and the 
curvature of the trajectories of a one-particle hamiltonian 
system. The formulas will involve solely the coordinates 
${\bf r}(x, y, z)$ and the velocities 
${\dot {\bf r}}({\dot x}, {\dot y}, {\dot z})$ of the particle.

We start from the well-known formulas [26] 
$$
\tau = - {{({\dot {\bf r}} \wedge {\ddot {\bf r}}) \, . \, \r3} 
          \over 
{\vert {\dot {\bf r}} \wedge {\ddot{\bf r}} \vert^2}} \qquad \quad k = {
{\vert {\dot {\bf r}} \wedge {\ddot{\bf r}} \vert} \over 
{\vert {\dot {\bf r}} \vert^3}}
\eqno (A1)
$$
for the torsion $\tau$ and the curvature $k$ of a curve (in \grt) 
in an arbitrary parametrization ${\bf r} \equiv {\bf r}(t)$. In 
the case where this curve is the trajectory of a particle (of 
mass $\mu$), we can use the Hamilton equations
$$
{\dot q} =   {\partial H \over \partial p} \qquad \quad
{\dot p} = - {\partial H \over \partial q} 
\eqno (A2)
$$
in order to express ${\ddot {\bf r}}$ and 
                    $\r3$ in 
(A1) in terms of the Hamilton function $H \equiv H(p,q,t)$. We shall 
employ the variable $p$ and $q$ in the Cartesian form 
$\left ( p_x = \mu {\dot x}, p_y = \mu {\dot y}, p_z = \mu {\dot z} \right )$
and $(x, y, z)$, respectively. Then, we have 
$$
{\ddot x} = - {1 \over \mu} {\partial H \over \partial x} \qquad 
\quad 
{\ddot y} = - {1 \over \mu} {\partial H \over \partial y} \qquad 
\quad 
{\ddot z} = - {1 \over \mu} {\partial H \over \partial z} 
\eqno (A3)
$$
and
$$
\x3 = - {1 \over \mu} \left( 
  {\partial^2 H \over \partial x^2}            {\dot x}
+ {\partial^2 H \over \partial y \partial x} {\dot y}
+ {\partial^2 H \over \partial z \partial x} {\dot z}
- {\partial^2 H \over \partial p_x \partial x} {\partial H \over \partial x}
- {\partial^2 H \over \partial p_y \partial x} {\partial H \over \partial y}
- {\partial^2 H \over \partial p_z \partial x} {\partial H \over \partial z}
+ {\partial^2 H \over \partial t \partial x}
\right)
\eqno (A4)
$$
with similar expressions for $\y3$ and $\z3$. As a result, the 
quantities $\tau$ and $k$ can be developed in terms of 
(${\dot x}, {\dot y}, {\dot z}$) and of the derivatives of $H$ 
which turn out to be functions of ($x, y, z$), 
(${\dot x}, {\dot y}, {\dot z}$) and $t$. This leads to formulas, not 
reported here, which are intricate but very easy to handle 
with a symbolic computer language like {\it Reduce} or {\it 
Macsyma}.

We now restrict ourselves to the case of conservative systems 
for which
$$
H = {1 \over 2 \mu} \left( {p_x}^2 + {p_y}^2 + {p_z}^2 \right) 
  + V(x,y,z). 
\eqno (A5)
$$
Then, the expressions for
$$
\tau = - {{\rm NUM} \over {\rm DEN}} \qquad \quad 
k = {\sqrt{\rm DEN} \over {\rm den}} \qquad \quad 
{\rm den} = \left( {\dot x}^2 + {\dot y}^2 + {\dot z}^2 \right)^{3/2} 
\eqno (A6)
$$
can be given a simple form. Indeed, we get
$$
\eqalign{
\mu^2 \times {\rm NUM} & = 
\left( {\dot y} {{\partial V} \over {\partial z}} 
     - {\dot z} {{\partial V} \over {\partial y}} \right) 
\left( {{\partial^2 V} \over {\partial x^2         }} {\dot x} + 
       {{\partial^2 V} \over {\partial y \partial x}} {\dot y} + 
       {{\partial^2 V} \over {\partial z \partial x}} {\dot z} \right) \cr 
                       & + 
\left( {\dot z} {{\partial V} \over {\partial x}} 
     - {\dot x} {{\partial V} \over {\partial z}} \right) 
\left( {{\partial^2 V} \over {\partial x \partial y}} {\dot x} + 
       {{\partial^2 V} \over {\partial y^2         }} {\dot y} + 
       {{\partial^2 V} \over {\partial z \partial y}} {\dot z} \right) \cr 
                       & + 
\left( {\dot x} {{\partial V} \over {\partial y}} 
     - {\dot y} {{\partial V} \over {\partial x}} \right) 
\left( {{\partial^2 V} \over {\partial x \partial z}} {\dot x} + 
       {{\partial^2 V} \over {\partial y \partial z}} {\dot y} + 
       {{\partial^2 V} \over {\partial z^2         }} {\dot z} \right) \cr 
\mu^2 \times {\rm DEN} & = 
\left( {\dot y} {{\partial V} \over {\partial z}} 
     - {\dot z} {{\partial V} \over {\partial y}} \right)^2 +
\left( {\dot z} {{\partial V} \over {\partial x}} 
     - {\dot x} {{\partial V} \over {\partial z}} \right)^2 +
\left( {\dot x} {{\partial V} \over {\partial y}} 
     - {\dot y} {{\partial V} \over {\partial x}} \right)^2 \cr 
}
\eqno (A7)
$$
or in compact form
$$
{\rm NUM} = {1 \over {\mu^2}} 
\left( {\dot {\bf r}} \wedge {\nabla V} \right) . 
\left( 
 ({\dot {\bf r}} \, . \, {\nabla {{\partial V} \over {\partial 
x}}}) \, {\bf i} + 
 ({\dot {\bf r}} \, . \, {\nabla {{\partial V} \over {\partial 
y}}}) \, {\bf j} + 
 ({\dot {\bf r}} \, . \, {\nabla {{\partial V} \over {\partial 
z}}}) \, {\bf k}
\right) \quad 
{\rm DEN} = {1 \over {\mu^2}} 
\vert {\dot {\bf r}} \wedge {\nabla V} \vert^2 
\eqno (A8)
$$
so that $\tau$ and $k$ are given by closed form expressions 
involving only 
($x, y, z$) and (${\dot x}, {\dot y}, {\dot z}$). Equations 
(A6) and (A7) are applied to $V = U_Q$ and $V = V_Q$ in the 
main body of this paper.
\bigskip
\smallskip

\centerline {\bf Acknowledgments}
\bigskip

This work has been achieved  in the framework of the 
France-Qu\'ebec  exchange programme   (project no 20~02~20~89). 
The kind hospitality extended to one of the authors (M.~K.) 
at the Centre de Recherches 
Math\'ematiques de l'Universit\'e de Montr\'eal on the occasion 
of several stays during the work on this project  is gratefully 
acknowledged.      The research of one of the authors (P.~W.) 
is partially supported by 
research grants from NSERC of Canada and FCAR of Qu\'ebec. Two 
of the authors (M.~K. and P.~W.) 
are indebted to Dr.~N.~W.~Evans from the Queen Mary 
College (London) for sending them preprints on his work and to 
Professor Ya.~A.~Smorodinsky for an interesting discussion. 
\bigskip
\smallskip

\vfill\eject

\centerline {\bf Bibliography}
\bigskip
\baselineskip = 0.68 true cm

 \item{[1]} H.~Hartmann, in~: {\it Sitzungsberichte der 
wissenschaftlichen Gesellschaft der J.-W.~Goe\-the 
Universit\"at}, Frankfurt am Main 
{\bf 10}, 107 (1972)~; {\it ibid.} Theor.~Chim.~Acta {\bf 24}, 201 
(1972)~; H.~Hartmann, R.~Schuck, and J.~Radtke, 
Theor.~Chim.~Acta {\bf 42}, 1 (1976)~; 
D.~Schuch, {\it Master Thesis}, J.-W.~Goe\-the 
Universit\"at, Frankfurt am Main (1978)~; 
H.~Hartmann and D.~Schuch, Int. J. Quantum Chem. {\bf 18}, 125 
(1980). 

 \item{[2]} C.~Quesne, J.~Phys.~A~:~Math.~Gen.~{\bf 21}, 3093 (1988).

 \item{[3]} M.~Kibler and T.~N\'egadi, Int. J. Quantum Chem. 
{\bf 26}, 405 (1984)~; {\it ibid.} Croat. Chem. Acta {\bf 57}, 1509 (1984).

 \item{[4]} M.~V.~Carpio and A.~Inomata, in~: {\it Path Integrals 
from meV to MeV}, Eds. M.~C. Gutzwiller, A. Inomata, 
J.~R.~Klauder, and L.~Streit (Singapore~: World Scientific, 1986). 

 \item{[5]} I.~S\"okmen, Phys. Lett. {\bf 115A}, 249 (1986).

 \item{[6]} C.~C.~Gerry, Phys. Lett. {\bf 118A}, 445 (1986).

 \item{[7]} M.~Kibler and P.~Winternitz, J. Phys. A~: Math. Gen. 
{\bf 20}, 4097 (1987). 

 \item{[8]} L.~Chetouani, L.~Guechi, and T.~F.~Hammann, Phys. 
Lett. {\bf 125A}, 277 (1987).

 \item{[9]} I.~V.~Lutsenko, G.~S.~Pogosyan, A.~N.~Sissakyan, and 
V.~M.~Ter-Antonyan, Teor.~i Mat.~Fiz. {\bf 83}, 419 (1990). 

 \item{[10]} M.~V.~Carpio-Bernido and C.~C.~Bernido, Phys. Lett.
{\bf 134A}, 395 (1989)~; 
{\it ibid.} Phys. Lett. {\bf 137A}, 1 (1989)~; 
M.~V.~Carpio-Bernido, 
J.~Phys.~A~: Math.~Gen. {\bf 24}, 3013 (1991)~; {\it ibid.}
J.~Math.~Phys.~{\bf 32}, 1799 (1991). 

 \item{[11]} J.~M.~Cai and A.~Inomata, Phys.~Lett.~{\bf 141A}, 315 
(1989).

 \item{[12]} A.~N.~Sissakian {\it et al}., preprint 
P2-89-814, JINR, Dubna (1989).

 \item{[13]} H.~Boschi Filho and A.~N.~Vaidya, Phys. Lett. 
{\bf 145A}, 69  (1990)~; {\it ibid.}           Phys. Lett. 
{\bf 149A}, 336 (1990)~; H.~Boschi-Filho, M.~de Souza, and 
A.~N.~Vaidya, J.~Phys.~A~: Math.~Gen.~{\bf 24}, 4981 (1991). 

 \item{[14]} O.~F.~Gal'bert, Ya.~I.~Granovskii, and 
A.S.~Zhedanov, Phys.~Lett.~{\bf 153A}, 177 (1991). 

 \item{[15]} A.~Guha and S.~Mukherjee, J.~Math.~Phys.~{\bf 28}, 840 (1987). 

 \item{[16]} M.~Kibler and T.~N\'egadi, Phys. Lett. 
{\bf 124A}, 42 (1987)~; {\it ibid.} in~: {\it Proc. 17th Int. Colloq. on Group 
Theoretical Methods in Physics}, Eds. Y.~Saint-Aubin and L.~Vinet 
(Singapore~: World Scientific, 1989). 

 \item{[17]} I.~S\"okmen, Phys.~Lett.~{\bf 132A}, 65 (1988).

 \item{[18]} L.~Chetouani, L.~Guechi, and T.~F.~Hammann, 
J.~Math.~Phys.~{\bf 30}, 655 (1989).

 \item{[19]} H.~D.~Doebner and E.~Papp, Phys.~Lett.~{\bf 144A}, 423 
(1990).

 \item{[20]} M.~Kibler and P.~Winternitz, Phys. Lett. {\bf 
147A}, 338 (1990). 

 \item{[21]} P.~Winternitz, Ya.~A.~Smorodinski\u \i , M.~Uhl\'\i \v r, 
and I.~Fri\v s, Yad. Fiz. {\bf 4}, 625 (1966) 
[Sov.~J.~Nucl.~Phys.~{\bf 4}, 444 (1967)]~; see also~: I.~Fri\v s, 
V.~Mandrosov, Ya.~A.~Smorodinsky, M.~Uhl\'\i \v r, and P. Winternitz, 
Phys.~Lett.~{\bf 16}, 354 (1965).

 \item{[22]} A.~A.~Makarov, J.~A.~Smorodinsky, Kh.~Valiev, and 
P.~Winternitz, Nuovo Cimento A {\bf 52}, 1061 (1967). 

 \item{[23]} N.~W.~Evans, Phys. Lett. {\bf 147A}, 483 (1990)~; {\it ibid.} 
Phys.~Rev.~A {\bf 41}, 5666 (1990)~; {\it ibid.} 
J. Math. Phys. {\bf 31}, 600 (1990).

 \item{[24]} A.~Joseph, Int.~J.~Quantum Chem.~{\bf 1}, 535 
(1967)~; see also~: C.~A.~Coulson and A.~Joseph, 
Int.~J.~Quantum Chem.~{\bf 1}, 337 (1967). 

 \item{[25]} M.~Moshinsky, C.~Quesne, and G.~Loyola, 
Ann.~Physics {\bf 198}, 103 (1990). 

 \item{[26]} M.~P.~do Carmo, {\it Differential Geometry of Curves 
and Surfaces} (New Jersey~: Pren\-ti\-ce-Hall, 1976) 

\bye